\documentclass[aip,sd,amsmath,amssymb,reprint]{revtex4-1}

\usepackage{graphicx}
\usepackage{bm}
\usepackage{color}
\usepackage{ulem}

\begin{document}


\title{Anomalous energy shift of laterally confined two-dimensional excitons}

\author{Shota Ono}
\email{shota\_o@gifu-u.ac.jp}
\affiliation{Department of Electrical, Electronic and Computer Engineering, Gifu University, Gifu 501-1193, Japan}
\author{Tomohiro Ogura}
\affiliation{Department of Electrical, Electronic and Computer Engineering, Gifu University, Gifu 501-1193, Japan}

\begin{abstract}
We theoretically investigate the energy of the ground state exciton confined to two-dimensional (2D) monolayers with circular shape. Within an effective mass approach employing a nonlocal screening effect on the Coulomb potential energy, we demonstrate how the exciton energy is correlated with the radius of the circle, electron-hole reduced mass, and 2D susceptibility. In addition, we show that a dead layer around the circle edge, into which the electron-hole pair cannot penetrate, is necessary for understanding the energy shift recently observed in monolayer WSe$_2$ quantum dots. 
\end{abstract}

\maketitle


\section{Introduction}
Since the discovery of monolayer graphene, two-dimensional (2D) materials are of great interest in the field of condensed matters.\cite{ajayan} In particular, monolayer transition metal dichalcogenides (TMD), showing a direct band gap with a few eV, has recently been investigated both from a fundamental point of view as well as for their practical applications.\cite{radisavljevic,mak} Interestingly, the optical absorption in those 2D monolayers has proven the existence of anomalous excitons, where the hydrogenic Rydberg series for the exciton binding energy is no longer observed.\cite{he,chernikov,wang} This is due to the macroscopic screening by charged particles immersed in a 2D monolayer. The 2D dielectric function is described not by a constant. Instead, it depends on the wavenumber, yielding an effective Coulomb potential energy between electron and hole in a real space \cite{rytova,keldysh,cudazzo}
\begin{eqnarray}
 V_{\rm{2D}}(r) &=& - v_0
 \left[ H_0 \left( \frac{r}{r_0}\right) - Y_0 \left( \frac{r}{r_0}\right)\right],
 \label{eq:2D}
\end{eqnarray}
where $v_0 = e^2/(4 \alpha_{\rm{2D}})$, $e$ is the elementary charge, $\alpha_{\rm{2D}}$ is the 2D susceptibility, $r$ is the distance between the charged particles in the 2D monolayer, $H_0$ and $Y_0$ are the Struve function and the Bessel function of the second kind, respectively, and $r_0 = 2\pi \alpha_{\rm{2D}}$. This potential shows a logarithmic divergence and a monotonic decay of $\propto r^{-1}$ for $r\ll r_0$ and $r \gg r_0$, respectively. In contrast to the case of $V_{\rm{2D}} \propto r^{-1}$ (Ref.~\cite{yang}), no exact solution for a two-particle system interacting through $V_{\rm 2D}$ given in Eq.~(\ref{eq:2D}) has been found, while an approximate solution for the ground state exciton has been derived recently.\cite{ganchev,olsen} See also recent review in Ref.~\cite{wang_mod}

Recent advances in experimental techniques have enabled synthesis of 2D monolayer quantum dots (QDs) with controllable size down to a radius of a few tenths of angstrom, boosting the investigation of novel phenomena of laterally confined excitons. The photoluminescence (PL) of excitons confined to 2D monolayer QDs has shown blueshift of the PL peak as the size of QDs is decreased.\cite{jin,wei} Such a behavior of 2D excitons seems to be analogous to that of three-dimensional (3D) excitons confined to semiconductor microcrystals with spherical shape.\cite{kayanuma} For the latter, the exciton motion would be influenced significantly when the size of microcrystals is comparable to or smaller than the effective Bohr radius, which leads to an energy shift of excitons. However, given Eq.~(\ref{eq:2D}), it is not trivial how 2D exciton property is related to the lateral confinement realized in 2D monolayer QDs because the concept of the effective Bohr radius is no longer valid in 2D monolayers. 

The purpose of this paper is two-fold. On one side, within an effective mass approach (EMA) exploiting Eq.~(\ref{eq:2D}), we investigate a correlation between the energy of 2D excitons confined to QDs and the material parameters; $\alpha_{\rm 2D}$ and the reduced mass of the electron and hole. On the other side, we apply our theory to the PL experiments for WSe$_2$ QDs.\cite{jin} We find that the calculated energy shift is significantly underestimated: A large amount of dead layer around the circle boundary, into which the electron-hole pair cannot penetrate, is needed to understand the experimental data. This implies that there would be strong perturbation potential around the edge, such as defects and reconstructed chemical bonds. 

\begin{figure}[b]
\center
\includegraphics[scale=0.35,clip]{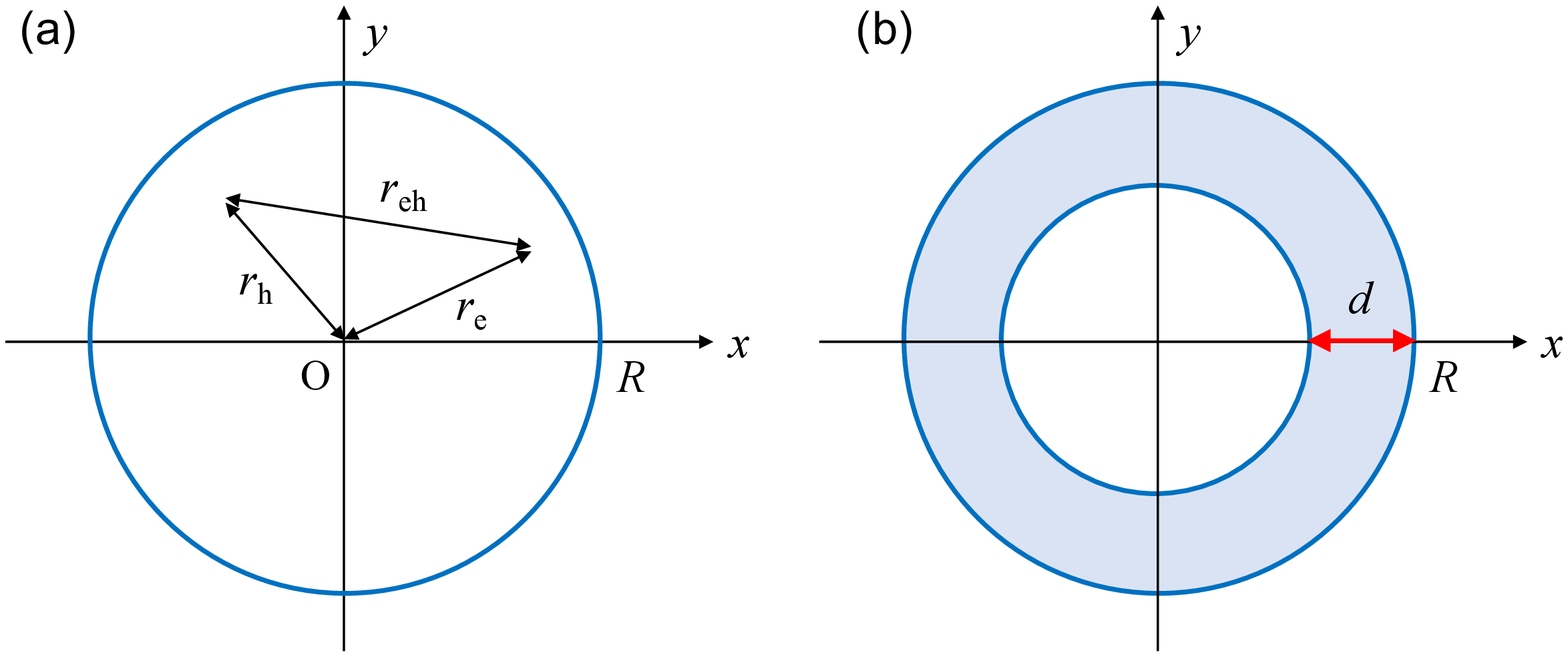}
\caption{\label{fig:coordinate} (a) Schematic illustration for $r_{\rm e}$, $r_{\rm h}$, and $r_{\rm eh}$ in a circle with a radius $R$. (b) The region of the dead layer having a width $d$. See Eq.~(\ref{eq:vc2}) for the definition. }
\end{figure}

\section{Formulation}
Let us first consider an electron and a hole inside a region with a circular shape: $\bm{r}_{i} = (r_i,\theta_i) \in {\cal A} = [0,R]\otimes[0,2\pi]$, where $\bm{r}_{i}$ denotes the position of the electron ($i={\rm e}$) and the hole ($i={\rm h}$) represented by polar coordinates $r_i$ and $\theta_i$. $R$ is the radius of the circle. We consider the two-particle Schr\"{o}dinger equation ${\cal H}\Psi = E \Psi$ with the Hamiltonian
\begin{eqnarray}
 {\cal H} &=& -\frac{\hbar^2}{2m_{\rm e}}\nabla_{\rm e}^{2} -\frac{\hbar^2}{2m_{\rm h}}\nabla_{\rm h}^{2} 
 + V_{\rm{2D}}(r_{\rm eh}) \nonumber\\
 &+& V_{\rm c}(r_{\rm e}) + V_{\rm c}(r_{\rm h}),
\end{eqnarray}
where $m_{\rm e}$ and $m_{\rm h}$ are the effective mass of electron and hole, respectively, and $r_{\rm eh} = \vert \bm{r}_{\rm e} - \bm{r}_{\rm h} \vert$. $V_{\rm c}(r_{i})$ is the confining potential: 
\begin{eqnarray}
 V_{\rm c}(r_{i}) =
 \begin{cases}
  0 \ \ \ {\rm for} \ \ \  r_i \le R \\
\infty \ \ \ {\rm for} \ \ \  r_i>R.
\end{cases}
\label{eq:vc1}
\end{eqnarray} 

To describe the motion of the electron and hole, we use the Hylleraas coordinates, where three variables, $r_{\rm e}$, $r_{\rm h}$, and $r_{\rm eh}$, are used (see Fig.~\ref{fig:coordinate}(a)). Assuming the cylindrical symmetry, the Hamiltonian for the ground state exciton is then expressed by 
\begin{eqnarray}
 {\cal H} &=& 
 -\frac{\hbar^2}{2m_{\rm e}}
 \left( \frac{\partial^2}{\partial r_{\rm e}^{2}} 
 + \frac{1}{r_{\rm e}}\frac{\partial}{\partial r_{\rm e}}
 + \frac{r_{\rm e}^{2} - r_{\rm h}^{2} + r_{\rm eh}^{2}}{r_{\rm e} r_{\rm eh}} 
 \frac{\partial^2}{\partial r_{\rm e}\partial r_{\rm eh}}
 \right)
 \nonumber\\
& & -\frac{\hbar^2}{2m_{\rm h}}
 \left( \frac{\partial^2}{\partial r_{\rm h}^{2}} 
 + \frac{1}{r_{\rm h}}\frac{\partial}{\partial r_{\rm h}}
 + \frac{r_{\rm h}^{2} - r_{\rm e}^{2} + r_{\rm eh}^{2}}{r_{\rm h} r_{\rm eh}}
  \frac{\partial^2}{\partial r_{\rm h}\partial r_{\rm eh}}
 \right)
  \nonumber\\
& & -\frac{\hbar^2}{2\mu}
 \left( \frac{\partial^2}{\partial r_{\rm eh}^{2}} 
 + \frac{1}{r_{\rm eh}}\frac{\partial}{\partial r_{\rm eh}}
 \right)
 + V_{\rm{2D}}(r_{\rm eh}) 
  \nonumber\\
 & & 
 +V_{\rm c}(r_{\rm e}) + V_{\rm c}(r_{\rm h}),
 \label{eq:hamiltonian}
\end{eqnarray}
where $\mu= m_{\rm e} m_{\rm h}/(m_{\rm e} + m_{\rm h})$ is the reduced mass of an electron and a hole. A variational method is used to compute $E(R) = \langle \Psi \vert {\cal H} \vert \Psi \rangle$ with 
\begin{eqnarray}
\langle \Psi \vert {\cal H} \vert \Psi \rangle
= \int_{0}^{R} dr_{\rm e} \int_{0}^{R} dr_{\rm h} 
\int_{\vert r_{\rm e} - r_{\rm h}\vert}^{r_{\rm e} + r_{\rm h}} dr_{\rm eh}
{\cal J}  (\Psi^{*} {\cal H}  \Psi)
\label{eq:braket}
\end{eqnarray}
and the Jacobian 
\begin{eqnarray}
{\cal J} = \frac{8\pi r_{\rm e} r_{\rm h} r_{\rm eh}}
{\sqrt{\left[ (r_{\rm e} + r_{\rm h})^2 - r_{\rm eh}^{2} \right] 
\left[r_{\rm eh}^{2} -  (r_{\rm e} - r_{\rm h})^2 \right]}}.
\end{eqnarray}
The trial function $\Psi$ includes the variational parameters $\{p\}=(p_1,p_2,\cdots)$. The minimization of $E(R)$ with respect to $\{p\}$ is done by the quasi-Newton method.\cite{recipe} We perform numerical integration for calculating Eq.~(\ref{eq:braket}), where the Boole's rule is used for the integrations with respect to $r_{\rm e}$ and $r_{\rm h}$ with 2000 grids, while the Chebyshev-Gauss quadrature \cite{cheb} is used for the integration of $r_{\rm eh}$ with 30 points. 

\section{2D monolayer}
Before studying the excitons in QDs, we construct a trial function for describing the exciton in a 2D monolayer with $R\rightarrow\infty$: The first, second, fifth, and sixth terms in the right hand side of Eq.~(\ref{eq:hamiltonian}) are set to be zero. $-E(\infty)$ can be regarded as the exciton binding energy $E_{\rm B}$. Given Eq.~(\ref{eq:2D}), the $1s$-type function is not an exact eigenfunction for the Hamiltonian of Eq.~(\ref{eq:hamiltonian}). We thus assume 
\begin{eqnarray}
\Psi (r_{\rm eh}) = N \left(1 + C X \right) e^{-r_{\rm eh} /a}, 
\label{eq:trial}
\end{eqnarray}
where $X=r_{\rm eh}/a_0$, $a_0=a_{\rm B}/\mu$, and $a_{\rm B}$ the Bohr radius of a hydrogen atom. $a$ and $C$ are the variational parameters: $a$ measures the spatial extent of the exciton wavefunction, while such an extent can be modulated by $C$. $N$ is the normalization factor.

\begin{figure}[t]
\center
\includegraphics[scale=0.45,clip]{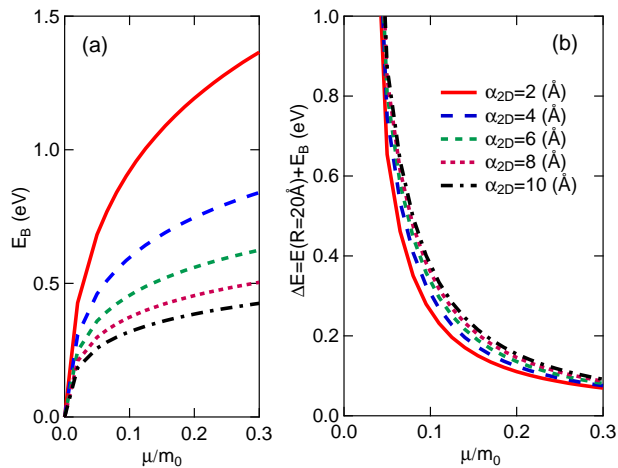}
\caption{\label{fig:bind} The $\mu$-dependence of (a) $E_{\rm B}$ and (b) the energy shift $\Delta E(R=20 {\rm \AA})$ from $E(\infty)=-E_{\rm B}$ for various $\alpha_{\rm 2D}$s. $m_0$ is the bare electron mass. }
\end{figure}

To rationalize the use of Eq.~(\ref{eq:trial}), we check the virial theorem. We define the energy expectation value with respect to the scaled wavefunction as
\begin{eqnarray}
 {\tilde E}
 &=& \int d\bm{r}_e\int d\bm{r}_h \Psi^{*} (\bm{r}_e, \bm{r}_h) 
 {\cal H}(L\bm{r}_e, L\bm{r}_h)\Psi (\bm{r}_e, \bm{r}_h) 
\end{eqnarray}
with $L$ being the scaling length. ${\tilde E}$ is minimum if the following relation is satisfied:\cite{ono}
\begin{eqnarray}
\frac{\partial {\tilde E}}{\partial L}\Big\vert_{L=1} = 0.
\end{eqnarray}
This gives the virial theorem $U/T = 2$ with 
\begin{eqnarray}
 T &=& \int dS \Psi^{*} (\bm{r}_{\rm eh}) 
\left(-\frac{\hbar^2}{2\mu}\nabla_{\rm eh}^{2} \right)
 \Psi (\bm{r}_{\rm eh}),
\\
 U &=& \int dS  \Psi^{*} (\bm{r}_{\rm eh}) 
 \left[ r_{\rm eh} \frac{d V_{\rm 2D}(r_{\rm eh})}{d r_{\rm eh}} \right]
 \Psi (\bm{r}_{\rm eh}),
\end{eqnarray}
where $dS=r_{\rm eh}dr_{\rm eh}d\theta_{\rm eh}$, $\bm{r}_{\rm eh} = (r_{\rm eh},\theta_{\rm eh})$, $\nabla_{\rm eh}^{2}=\partial^2/\partial r_{\rm eh}^{2} + (1/r_{\rm eh})\partial/\partial r_{\rm eh}$, and 
\begin{eqnarray}
\frac{d V_{\rm 2D}(r_{\rm eh})}{d r_{\rm eh}} 
= - \frac{v_0}{r_0}
\left[ \frac{2}{\pi } - H_{1} \left(\frac{r_{\rm eh}}{r_0}\right) 
+ Y_{1} \left(\frac{r_{\rm eh}}{r_0}\right) \right].
\nonumber\\
\end{eqnarray}
The use of Eq.~(\ref{eq:trial}) gives a ratio of $U/T=1.998\pm 0.001$ for infinite 2D monolayers. Note that the relation $U/T = 2$ is satisfied only when $R\rightarrow\infty$. When $R$ is finite, the relation $U/T = 2$ may not be satisfied due to the presence of the infinite barrier potentials $V_{\rm c}(r_{\rm e})$ and $V_{\rm c}(r_{\rm h})$.

The magnitude of $E_{\rm B}$ depends on the material parameters in Eq.~(\ref{eq:hamiltonian}), $m_{\rm e}$, $m_{\rm h}$, and $\alpha_{\rm 2D}$. Below, we assume $m_{\rm e}=m_{\rm h}=2\mu$. This is reasonable assumption for many TMDs around the band edges at K point. Figure \ref{fig:bind}(a) shows the $\mu$-dependence of $E_{\rm B}$ for various $\alpha_{\rm 2D}$s. The curve of $E_{\rm B}$-$\mu$ is not linear, which is different from the case of the standard Coulomb potential, $E_{\rm B}\propto \mu$ (Ref.~\cite{AM}). As $\alpha_{\rm 2D}$ increases, $E_{\rm B}$ approaches zero slowly. These properties originate from the fact that the screening is nonlocal in real space.\cite{olsen} Note that if we set $C=0$ in Eq.~(\ref{eq:trial}), $E_{\rm B}$ is underestimated by a few percent. This indicates that although the exciton Bohr radius cannot be defined exactly when Eq.~(\ref{eq:2D}) is used, the value of $a$ determines the spatial extent of excitons in the first approximation.

\section{2D monolayer QD}
Since Eq.~(\ref{eq:trial}) is a good function for the 2D exciton, we define the trial function in a 2D monolayer QD with a circular shape as 
\begin{eqnarray}
\Psi (r_{\rm e},r_{\rm h},r_{\rm eh}) &=& 
N J_0 \left(\frac{\gamma_{01}r_{\rm e}}{R} \right) 
J_0 \left(\frac{\gamma_{01}r_{\rm h}}{R} \right) 
\nonumber\\
&\times&
\left(1 + C X \right) e^{-r_{\rm eh} /a}, 
\label{eq:trial2}
\end{eqnarray}
where $J_0$ is the zeroth order Bessel function of the first kind and $\gamma_{ij}$ is the $j$th zero-point of the $i$th order Bessel function. The product $J_0 \left(\gamma_{01}r_{\rm e}/R \right) J_0 \left(\gamma_{01}r_{\rm h}/R \right) $ and the factor $(1 + C X )\exp\left(-r_{\rm eh} /a\right)$ in Eq.~(\ref{eq:trial}) describe the uncorrelated part and the correlation between electron and hole, respectively. Similar form of trial function has been proposed for excitons in 3D and 2D nanostructures,\cite{kayanuma,semina} while another form of exciton function can be found in the study of graphene QDs.\cite{li}

Generally, the exciton with small $E_{\rm B}$ has a large exciton radius. Intuitively, when the size of QDs is smaller than or comparable to the effective exciton radius in an infinite 2D monolayer, the motion of the exciton is severely affected by the confinement potential, yielding a significant shift of the exciton energy. This is illustrated in Fig.~\ref{fig:bind}(b): The energy shift $\Delta E = E(R=20{\rm \AA})-E(\infty)$ as a function of $\mu$ for various $\alpha_{\rm 2D}$. The smaller $\mu$ (the larger $\alpha_{\rm 2D}$) is, the larger $\Delta E$ becomes. 

\begin{figure}[t]
\center
\includegraphics[scale=0.45,clip]{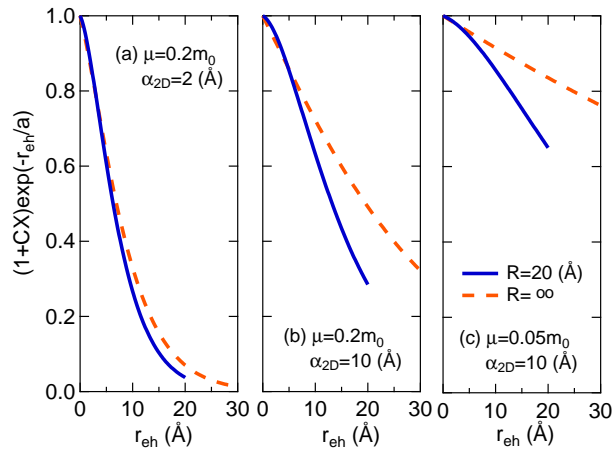}
\caption{\label{fig:wave} Exciton wavefunction, $(1+CX){\rm exp}(-r_{\rm eh}/a)$ given in Eqs.~(\ref{eq:trial}) and (\ref{eq:trial2}), for the cases of $R=\infty$ and $R=20$ \AA. (a) $\mu=0.2m_0$ and $\alpha_{\rm 2D}=2$ \AA, (b) $\mu=0.2m_0$ and $\alpha_{\rm 2D}=10$ \AA, and (c) $\mu=0.05m_0$ and $\alpha_{\rm 2D}=10$ \AA. }
\end{figure}

We next investigate how the exciton wavefunction is deformed when confined to 2D monolayer QDs. Figure~\ref{fig:wave} shows the exciton part of the wavefunction, $(1+CX){\rm exp}(-r_{\rm eh}/a)$ given in Eqs.~(\ref{eq:trial}) and (\ref{eq:trial2}), for the cases of $R\rightarrow\infty$ and $R=20$ \AA. When $\mu=0.2m_0$ and $\alpha_{\rm 2D}=2$ \AA \ [see Fig.~\ref{fig:wave}(a)], the exciton size is less than 20 \AA \ in the case of $R\rightarrow\infty$ (dashed). This leads to a small deformation of the wavefunction in its tail (solid) when $R$ is decreased to 20 \AA. As $\alpha_{\rm 2D}$ increases or $\mu$ decreases, the exciton wavefunction becomes quite sensitive to the size of QDs, where largely diffuse wavefunction realized in $R\rightarrow\infty$ is changed into localized wavefunction when $R=20$ \AA, as shown in shown in Figs.~\ref{fig:wave}(b) and \ref{fig:wave}(c). The smallness of the exciton size gives rise to an increase in the kinetic energy and thus leads to an increase in $\Delta E$, as demonstrated in Fig.~\ref{fig:bind}(b). 

\begin{figure}[t]
\center
\includegraphics[scale=0.45,clip]{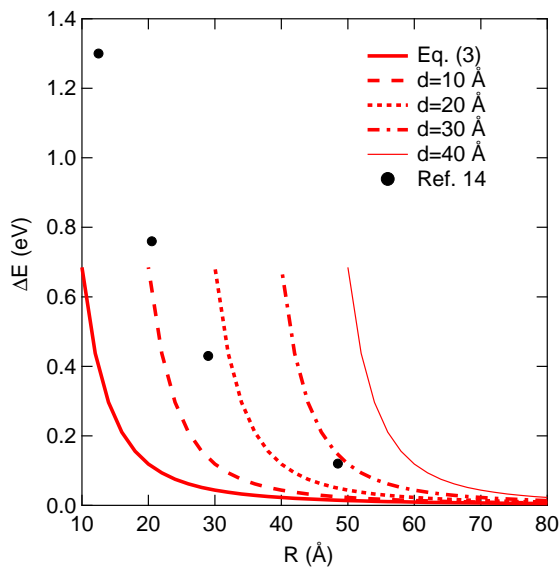}
\caption{\label{fig:exp} The energy shift $\Delta E$ as a function of $R$ for WSe$_2$ with the use of Eqs.~(\ref{eq:vc1}) and (\ref{eq:vc2}). For the latter, the curves for several $d$s are plotted. The experimental data of Ref.~\cite{jin} are also plotted. }
\end{figure}

\section{Application to experiment}
Let us apply the present model to interpret the PL of the monolayer QDs. We focus on the experiment of WSe$_2$ QDs\cite{jin}: The magnitude of the blueshift is 1.3 eV when $R$ is decreased to 12.5 \AA. The material parameters used in the present study are $\mu=0.23m_0$ and $\alpha_{\rm 2D}=7.35$ \AA \ suggested in Ref.~\cite{olsen}, yielding $E_{\rm B}=0.5$ eV. As shown in Fig.~\ref{fig:exp}, the energy shift calculated by using Eq.~(\ref{eq:vc1}) (solid) is significantly underestimated compared to the experimental data. This may imply that the boundary condition imposed is not suitable for modeling the realistic situation. To remedy it, we introduce a {\it dead layer} around the circle edge, into which the electron and hole cannot penetrate. This is expressed by 
\begin{eqnarray}
 V_{\rm c}(r_{i}) =
 \begin{cases}
  0 \ \ \ {\rm for} \ \ \  r_i \le R-d \\
\infty \ \ \ {\rm for} \ \ \  r_i>R-d,
\end{cases}
\label{eq:vc2}
\end{eqnarray} 
where $d$ is the width of the dead layer (see also Fig.~\ref{fig:coordinate}(b)). In Fig.~\ref{fig:exp}, $\Delta E$-$R$ curves for several $d$s are also plotted. The calculated curve with $d = 20$ \AA \ approximately reproduces the experimental data for $R=29$ and $48.5$ \AA, while a significant deviation is observed for $R\ll 30$ \AA. 

We discuss the physical meaning of the magnitude of $d \simeq 20$ \AA, whose width is about six times larger than the lattice constant of WSe$_2$. First, the shape of the realistic QDs is not a complete circle due to the discreteness of atoms. Second, a reconstruction of chemical bonds would occur around the circle edge, which gives rise to the change in the onsite and hopping energies in a sense of the tight-binding (TB) approximation. These perturbations are strong enough to affect the excitonic properties as well as the quasiparticle properties of nanoscale QDs. Within the EMA, such effects should be modeled by a large amount of the dead layer above.

Note that if we set $\mu=0.06m_0$ and $\alpha_{\rm 2D}=40$ \AA, the calculated energy shift can reproduce the experimental observation. However, with such parameters, $E_{\rm B}$ is estimated to be $0.1$ eV, which is much smaller than a typical value of $E_{\rm B}$ for TMDs; For example, $E_{\rm B}=$0.63 eV in MoS$_2$.\cite{qiu2} The small value of $\mu$ for reproducing the experimental data may imply that a Dirac character\cite{trushin} of the quasiparticle band is important to understand the 2D exciton properties correctly. Nevertheless, such a treatment would not solve the present problem fundamentally since it is also based on the continuum model. The use of an atomistic approach is then highly desired. For example, TB method is useful to study the excitonic properties in nanostructures. Ozfidan {\it et al}. have studied the optical properties of colloidal graphene QDs with the combined uses of TB, Hartree-Fock, and configuration interaction approaches.\cite{ozfidan} They have shown that the computed optical spectra are in agreement with experimental data. We expect that similar theoretical approaches would resolve the anomalous energy shift in TMDs.

\section{Summary}
We have studied the property of 2D excitons confined to QDs within the standard EMA, where the infinite barrier potentials are imposed to the electron and hole outside a circle. The exciton energy increases with decreasing the radius of the circle, while the magnitude of the energy shift strongly depends on the reduced mass and 2D susceptibility. We also calculated the exciton energy shift of WSe$_2$ QDs and found that the magnitude of the energy shift is significantly underestimated compared to the experiment. The finite width of dead layer around the circle edge is needed to reproduce the experimental data, implying that there are defects and reconstructed chemical bonds around the edge. We expect that the present work will stimulate further investigations on the optical properties of 2D TMDs with various geometries both theoretically and experimentally. 

\begin{acknowledgments}
We acknowledge helpful discussions with M. Aoki. This study is supported by a Grant-in-Aid for Young Scientists B (No. 15K17435) from JSPS.
\end{acknowledgments}


\end{document}